
\magnification=1200
\def\ni{\noindent}
\def\.{\mathaccent 95}
\def\a{\alpha}
\def\be{\beta}

\def\de{\delta}

\def\la{\lambda}

\def\om{\omega}
\def\Ga{\Gamma}

\def\La{\Lambda}

\def\Om{\Omega}

\def\frac#1#2{{\textstyle{{#1}\over {#2}}}}
\def\ni{\noindent}
\def\lsim{\mathrel{\rlap{\lower4pt\hbox{\hskip1pt$\sim$}}
    \raise1pt\hbox{$<$}}}
\def\gsim{\mathrel{\rlap{\lower4pt\hbox{\hskip1pt$\sim$}}
    \raise1pt\hbox{$>$}}}
\def\sqr#1#2{{\vcenter{\vbox{\hrule height.#2pt
         \hbox{\vrule width.#2pt height#1pt \kern#1pt
         \vrule width.#2pt}
         \hrule height.#2pt}}}}

\newbox\grsign \setbox\grsign=\hbox{$>$} \newdimen\grdimen \grdimen=\ht\grsign
\newbox\simlessbox \newbox\simgreatbox
\setbox\simgreatbox=\hbox{\raise.5ex\hbox{$>$}\llap
     {\lower.5ex\hbox{$\sim$}}}\ht1=\grdimen\dp1=0pt
\setbox\simlessbox=\hbox{\raise.5ex\hbox{$<$}\llap
     {\lower.5ex\hbox{$\sim$}}}\ht2=\grdimen\dp2=0pt

%
%

\def\ref#1  {\noindent \hangindent=24.0pt \hangafter=1 {#1} \par}

\def\doublespace {\smallskipamount=6pt plus2pt minus2pt
                  \medskipamount=12pt plus4pt minus4pt
                  \bigskipamount=24pt plus8pt minus8pt
                  \normalbaselineskip=24pt plus0pt minus0pt
                  \normallineskip=2pt
                  \normallineskiplimit=0pt
                  \jot=6pt
                  {\def\smallskip {\vskip\smallskipamount}}
                  {\def\medskip   {\vskip\medskipamount}}
                  {\def\bigskip   {\vskip\bigskipamount}}
                  {\setbox\strutbox=\hbox{\vrule 
                    height17.0pt depth7.0pt width 0pt}}
                  \parskip 12.0pt
                  \normalbaselines}

\font\gkvec=cmmib10                         
\def\bomega{\hbox{{\gkvec\char33}}}                  

\def\lb{\langle}
\def\rb{\rangle}

\def\bw{\bar{\omega}}
\def\bv{\bar v}
\def\ts{\times}
\def\lb{\langle}
\def\rb{\rangle}
\def\curl{\nabla {\ts}}

\def\bfvp{{\bf v}'}
\def\bfjp{{\bf j}'}

\def\bfwp{{\bomega}'}
\def\bfbp{{\bf b}'}

\def\nb{\nabla}
\def\curl{\nb\ts}

\def\b0{b'^{(0)}}
\def\v0{v'^{(0)}}
\def\w0{\omega'^{(0)}}
\def\bb0{\bfbp^{(0)}}
\def\bv0{\bfvp^{(0)}}
\def\bw0{\bfwp^{(0)}}
\def\bj0{\bfjp^{(0)}}

\def\ni{\noindent}

\centerline{\bf On Proton Energization in Accretion Flows}
\medskip
\centerline {Eric G. Blackman}
\centerline {Institute of Astronomy, Madingley Road, 
Cambridge, CB3 OHA, UK}
\centerline{(submitted to MNRAS)}
\doublespace
\ni {\bf ABSTRACT}:  Two-temperature advection dominated accretion flow (ADAF) 
or hot ion tori (HIT) models help explain 
low luminosity stellar and galactic accreting sources and may complement 
observational support for black holes in nature.  But low radiative 
efficiencies demand that ions receive a fraction $\eta \gsim 99\%$ of energy 
dissipated in the turbulent accretion. The $\eta$ depends on the ratio of 
particle to magnetic pressure. If compressive modes of dissipation, like 
magnetic mirroring, dominate incompressible modes, then even when the pressure
ratio is O(1), the required large $\eta$ can be attained.
However the relative importance of compressive vs. incompressible modes is
hard to estimate.  The larger up in the turbulent cascade, the more 
compressible the turbulence.  The relevant length 
scale for particle energization can be determined by equating the dominant 
eddy turnover time to the time for which an energy equal to that in the 
turbulence can be drained.  Based on the large scales 
resulting from this estimate, it is suggested that compressive mirroring 
may be important.  Also, regardless of the precise $\eta$ or dissipation 
mechanism, non-thermal protons seem natural in two-
temperature discs because all dissipation mechanisms, and the use of an 
isotopic pressure, require wave-particle resonances that operate only on a 
subset of the particles. Finally, it is briefly mentioned how
mirroring may help to generate an ADAF or HIT in the first place.

\ni Key words: accretion, accretion discs; acceleration of particles; 
turbulence; galaxies: general; binaries: general;  Galaxy: centre 
\vfill
\eject

\centerline{\bf 1. Introduction}

Magnetized accretion discs have become the most convincing physical paradigm 
to explain emission from the central engines of active galactic nuclei (AGN) 
and X-ray binary sources (Frank et al. 1992).  
The observed radiation comes from the energy dissipation required to 
maintain steady accretion of material
onto the central object.  As molecular viscosity is incapable of 
providing the required accretion rates, turbulent viscosity
is necessary. For thin discs, this can be generated
by shear and magnetic fields (Balbus \& Hawley 1991).  For thick
discs, something similar may occur, though in this case
angular momentum transport may ultimately require a global approach.

Nevertheless, in complement to thin disc solutions for sources requiring high
radiation efficiency accretion (e.g. Frank et al. 1991), two-temperature thick advection dominated flows (ADAFs) or hot ion tori (HIT) 
(e.g. Ichimaru 1977; Paczynski \& Bitsnovatyi-Kogan 1981; Rees et al. 1982; Narayan \& Yi 1995) have received much attention 
in an effort to explain sources requiring a low radiation efficiency.  
Here the ions are assumed to receive the
energy dissipated by the steady accretion without having enough time
to transfer their energy to the cooler electrons before falling onto the
central object.  Some or most of the dissipated energy is advected, not
radiated, as it would have been if electrons received all of the dissipated 
energy.  Such models have been at least partially 
successful in explaining quiescent galactic centres
(Rees 1982; Narayan et al. 1995; Mahadevan 1998; Fabian et al. 1995, but see
 DiMatteo et al. 1998) 
and stellar X-ray binary systems (Narayan et al. 1996) with radiative
efficiencies $\le$ 1/100 that of  thin disc solutions.
When the central object is a black hole, the advected energy is lost forever
rather than re-radiated as it would be for a neutron star.
Precisely such observed differences between corresponding X-ray binary systems
have been purported to provide evidence for  
black hole horizons (Narayan et. al 1997). 

There have been only a handful of other papers addressing how the viscous 
dissipation might energize particles in accretion flows 
(Gruzinov 1997; Quataert 1997; Quataert \& Gruzinov 1998) 
and none addressing thermal vs. non-thermal 
particle distributions. Both of these issues are 
extremely important for ADAF type models because: 1) a two-temperature solution
is not sufficient to explain a low radiation efficiency and 2) 
interpretation of observations of the Galactic centre 
suggest that the protons are non-thermal when an ADAF
model is employed (Mahadevan 1998).
The potential catastrophe for ADAF/HIT models, if electrons are preferentially energized over protons, was partially explored in 
(Bitsnovatyi-Kogan \& Lovelace 1997).  
Because electrons cool much faster than  ions, even
if 1/2 of the dissipated energy went into electrons, 
a two-temperature solution would still 
result.  But in this case, 50\% of the dissipated energy would
be radiated---far too much to explain low luminosity sources.
More explicitly, I define $q_t$ 
as the magnitude of the energy density input
rate into particles.  Then $q_t=q_p^+ + q_e^+=\eta q_t +(1-\eta)q_t$
where $q_p^+$ and $q_e^+$ are the magnitudes of energy input rate 
into protons and electrons, and
 $\eta$ is the fraction of $q_t$ that goes into protons.
In the steady state, energy loss rates are equal to energy
gain rates so that  when advection is included,  we have for the protons
$q_p^+=\eta q_t= q_{a}^- - q_{pe}
=f\eta q_t+(1-f)\eta q_t,$
where $q_{a}=f\eta q_t$ is the rate associated with advection, 
$q_{pe}$ is the rate of transfer from ions to electrons, and
$f$ is the fraction of the proton energy loss rate associated with advection.
For the electrons, we thus have
$q_e^+=(1-\eta) q_t +q_{pe}= q_t - q_{a}= q_t(1-f\eta).$
Since $q_e^+=q_e^-$, where $q_e^-$ is the luminosity density, 
the quantity $(1-f\eta)$ must
be $<<1$ to explain quiescent sources. 
Standard treatments (e.g. Rees et al. 1982; Narayan \& Yi 1995)
assume $\eta=1$ so that
$1\%$ radiative efficiency would correspond to $f=0.99$.
The important questions are: (A) When can $\eta \gsim 0.99 $ be justified? 
(B) Are protons non-thermal? and
(C) Is there a faster than Coulomb coupling between electrons
and protons (e.g. Begelman \& Chiueh 1988) that destroys the ADAF solution?  
I will address (A) and (B) here.

Proton energization by incompressible 
(Quataert 1997; Gruzinov 1997; Quataert \& Gruzinov 1998)
and/or compressive modes of dissipation (discussed herein) both depend 
on the ratio of particle to magnetic pressure.
Section 2 addresses the relation between magnetic, turbulent, and 
particle energy densities in ADAFs, relating them to the viscosity parameter
$\a$.  Section 3 discusses the threat of electron runaway.  
Section 4 employs a very physical approach to acceleration by 
magnetic mirroring and derives the time scale for particle energy 
doubling for two distinct limits of the average particle speed.  Section 5 
discusses energization in ADAFs, first addressing why protons 
are likely to be non-thermal, regardless of the acceleration mode.
The mirroring results are then specifically applied to ADAFs and 
the scale in the turbulent cascade where mirroring is favored is estimated.
Because larger scales are significantly compressive, and the
resulting derived scale can be large, mirroring may be important.
Magnetic mirroring type processes can favor protons to the 
extent required for a wider range of average particle to magnetic
pressure ratios than found by Quataert \& Gruzinov (1998)
from dissipation of incompressible Alfv\'en waves, but the
relative fractions of incompressible vs. compressible modes of dissipation
are hard to determine. 
The possibility that mirroring may help provide a thermal instability
which initially forms an ADAF is briefly addressed.  

\centerline{\bf 2. Relation Between Viscosity Parameter and Pressure Ratio}

The standard parameterization of accretion disc turbulent viscosity 
for thin discs is (Shakura \& Sunayev 1973) 
$$\nu_{T}=\a C_s H\sim V_T l_T/3,\eqno(1)$$ 
where $\a$ is the viscosity parameter and $C_s$ and $H$ are the sound speed
and disc height, and $V_T$ and $l_T$
are the outer (i.e. dominant energy containing)
random (turbulent) flow speed and scale.  
%
For thin discs, the scale of the turbulence is always much less
than the radius of the disc.  
Turbulent kinetic and magnetic energy densities rapidly approach
equipartition from field line stretching.  
A magneto-rotational shearing instability (Balbus \& Hawley 1991) 
likely drives 3-D MHD turbulence.
In a steady state, dissipation of the turbulence into particles 
combats the symbiotic growth of magnetic and kinetic turbulent energy.
Both magnetic and kinetic turbulent energies 
incur a decaying power-law energy spectrum 
(like Kolmogorov (1941) or Kraichnan (1965))
with the largest scales of the turbulence containing the most energy.
Since the sound speed is constant on all scales, the largest
scales are thus the most compressive.  For thin discs,
the outer turbulent scale is significantly smaller than the disc radius,
but for standard ADAFs, there is not such a strong scale separation 
(Blackman 1998).

For thin discs, we can derive a 1-to-1 link between $\a$ and 
$\beta_p^{-1}\equiv V_A^2/C_s^2 \equiv 6(1-\beta_{a})$, where $V_A$ is the
Alfv\'en speed, and $\be_{a}$ is used in ADAF modeling. 
In the steady-state, the largest eddy turnover time 
$t_{T}=l_T/V_T$ must equal the shearing instability growth time
driving the turbulence, that is: $t_{T}\simeq R/V_{\phi}$, where
the rotation speed $V_{\phi}\sim V_{K}$, the Keplerian speed. 
 Since $V_A\sim V_{T}$
in the saturated state from field line stretching (e.g. Parker 1979; Blackman
1998) and $C_s/V_{K}= H/R$ from hydrostatic equilibrium, we have    
$\nu=\a C_s H\simeq V_{T}l_{T}/3 \sim V_A^2(R/3V_{\phi})=V_A^2 (H/3C_S)$ which implies that 
$$\a \simeq 2(1-\beta_{a})(V_{K}/V_\phi)\simeq 2(1-\beta_{a})
\eqno(2)$$ 
for thin discs.  This result basically agrees with numerical
simulation results (e.g. Stone et. al 1996).

For ADAFs, the ratio of $V_K/V_\phi$ can be so low that
(2) is inappropriate: in this case the resulting eddy scale
implied by the relation would be larger than $H\sim R$.
We can instead obtain an upper limit on $\a$ for ADAFs
that comes from the constraint 
$$l_T<H\sim R.\eqno(3)$$
Then from (1), the definition of $\be_{a}$, and 
$V_T=V_A$ from field line stretching,  we find
$$\a \le (2/3^{1/2})(1-\be_a)^{1/2}.\eqno(4)$$
showing that $\beta_{a}$ and $\a$ are not independent.



\centerline{\bf 3. On Electron Runaway}

Bistnoyati-Kogan \& Lovelace (1997) pose an interesting question:
why can't direct acceleration from electric fields drain energy into 
electrons, destroying ADAFs?  I address this here.
First, note the generalized Ohm's Law (e.g. Scudder 1986)
${\bf E}= -{\bf V}_e/c \ts {\bf B} +\sigma^{-1} {\bf J}- m_e({\bf V}_e\cdot\nabla{\bf V}_e)/e-\nabla P_e/(en_e)$
where $\bf B$ is the magnetic field, $\bf J$ is the current density, $P_e$ is 
the electron pressure, ${\bf V}_e$ is the bulk
electron velocity, $\sigma^{-1}$ is the resistivity, $m_e$ is the
electron mass, $n_e$ is the density, 
and $e$ is the electric charge.
For the plasmas of interest, 
a characteristic magnitude of  $\bf E$ parallel to $\bf B$ is  
given by the second last term, and using $V_e\lsim V_T\sim  V_A\lsim C_s$, 
we have
$|{E}_{||}| \sim k_bT_e/(\de l\ e)\sim 2\ts 10^{-14}(T_e/10^{9}{\rm K})(\de l/10^{13}
{\rm cm})^{-1}$, 
where $\de l$ is the gradient length, $T_e$ is the electron 
temperature and $k_b$ is the Boltzmann constant.

For $E_{||}$ to produce ER, it 
would have to exceed the Dreicer electric field (Dreicer 1962; Holman 1985; 
Bitsnovatyi-Kogan \& Lovelace 1997)
$E_D=e^2 ln {\Ga}/\la_d^2=1.8\ts 10^{-7}(ln\Ga/20)(n_e/10^{12}{\rm cm^{-3}})^{1/2}$ 
$(T_e/10^9{\rm K})^{-1}$St-Volt/cm, 
using $\la_D\sim 6.65 (T_e^{1/2}/n_e)$cm, for the Debye length.
Whether $E_{||} > E_D$ depends on the size of $\de l$.
For AGN, $E_D$ is only exceeded on scales $10^6$ times smaller than
the turbulent outer scale. For stellar size X-ray binary 
ADAF systems, the outer scale is $\sim 10^7$cm, so in principle
ER is possible throughout the flow.
But the accelerated electrons can never produce 
a current which induces a magnetic field in excess of the inferred ambient
field.  This gives  an upper limit (Holman 1985) to the size 
of field gradients that generate ER, namely
$\de l\le 8(B/10^{4}{\rm G})(n_e/10^{12}{\rm cm^{-3}})^{-1}(T_e/10^9{\rm K})^{-1/2}
(E_D/E_{||})$cm.  For all relevant accretion discs, this scale in the cascade 
is always way below that at which magnetic
mirroring, employed in the next section, could 
have already drained most of the 
energy in the cascade.  Nevertheless, if mirroring is not important, or
equivalently, 
if a significant component of the turbulence cascades to incompressible 
scales before draining,
then the cascade may proceed down to this scale where ER, or other
(e.g. Quataert \& Gruzinov 1998)
electron energization processes may be important.

However, the more extreme ER of  Bistnovati-Kogan \& Lovelace (1997)
is not likely. They employ the mean electric
field, obtained by coarsely averaging
$\bf E =\lb {\bf E} \rb+{\bf E}_T$ over the turbulent scales $l_T$.
The dominant terms in this mean Ohm's law are then
$\lb{\bf E}\rb\simeq -\lb{\bf V}_T\ts{\bf B}_T\rb -\lb{\bf V}\rb\ts\lb{\bf B}\rb$
where the turbulent EMF (Parker 1979) is, in kinematic theory,  
$\lb{\bf V}_T\ts {\bf B}_T\rb=(\a_d/c)\lb{\bf B}\rb-\beta_d\curl\lb{\bf B}\rb$
where $\a_d$ is a pseudo-scalar helicity, and $\beta_d$ is the turbulent diffusivity.
A representive magnitude of $\lb E\rb_{||}$ 
using $|\a_d|\sim V_T/3$, is $\lb E_{||}\rb\sim V_T \lb  B\rb /3c$.
Assuming that $\lb B \rb\sim B$, then
$\lb E_{||}\rb\sim 3\ts 10^3(V_T/10^{10}{\rm cm/sec})(B/10^4{\rm G})$St-Volt/cm.
Thus $\lb E\rb>>E_D$ and one might be tempted to conclude that
runaway electron acceleration is extreme. 
But particles do not actually see $\lb{\bf E}\rb$, since the 
average is only defined on scales larger than $l_T$.  Thus extreme
ER should not occur.  

\centerline {\bf 4.  Energization by Magnetic Mirroring}

{\bf 4.1 Basic Physical Picture}

Fermi energization, or magnetic mirroring 
of particles off magnetic compressions (Fermi 1949; Spitzer 1962) provides a 
means of dissipation of compressive turbulent energy into particles.
Condsider a field, $B_T$, which represents the field on the largest
turbulent scale, superimposed on which is  a smaller scale 
compression, so the total field in the compression is $B= 
B_T+\de B$.  
These compressions travel along field lines at speeds $\sim V_A$,  
and can transfer energy to particles: 
consider what happens as the particle traveling along $B_T$ 
interacts with the compression $B_T + \de B$.  
Since the magnetic force is perpendicular to the particle velocity, 
as long as the magnetic gradient scale $>>$ particle 
gyro-radius (adiabatic approximation), the particle's angular momentum 
and energy are conserved in the frame of a magnetic compression at rest.
Denoting quantities in this frame by a prime and working 
in the  non-relativistic limit, the energy and angular momentum magnitudes 
are given by $u'=m v'^2/2$ and
$j'=m v_{\perp}'r_g=m^2 c {v_\perp}'^2/(e B)$, where $m$ is 
the particle mass, $v',v'_\perp$ are the total speed and speed 
perpendicular to the field  
and $r_g$ is the gyro-radius. The constancy of both $u'$ and $j'$
implies that $v_\perp'^2/B=v'^2 sin^2 \phi'/B$ is also constant.
Thus, because $v'$ is constant, $sin^2\phi'\propto B=B_T+\de B$,
or 
$$sin^2\phi'=sin^2\phi_{T}'(B_T+\de B)/B_T.\eqno(5)$$  
When $sin\phi'=1$, the particle reflects.  Thus
there exists a minimum pitch angle the particle
must have with respect to the ambient $B_T$ such that 
it can reflect upon entering the compression.  This is given by
$$sin^2 \phi_{T,min}'\equiv sin^2 \phi_{min}'=B_T/(B_T+\de B) \eqno(6)$$
(see also figure 1).  In the lab frame, the moving compression then 
boosts the velocity component of a given
particle parallel to $B_T$.  For energy to be gained
from repeated reflections, the boost must be 
rapidly isotropized by particle generated waves 
(Eilek \& Hughes 1991; Larosa et al. 1996) as discussed further in section 5. 
Assuming isotropy in the lab frame, 
in the frame of a moving magnetic fluctuation the velocity distribution is 
centered around $\sim V_A$. 
The minimum angle for mirroring by a magnetic compression then
gives a minimum speed that particles need to reflect: 
$$v_{min}=V_A sin \phi'_{min}=V_A B_T^{1/2}/(B_T+\de B)^{1/2}\ \sim V_A
\ {\rm for}\  \de B/B_T< 1,\eqno(7)$$ 
as seen in figure 1.


{\bf 4.2 Time Scale for Energy Doubling}

Different regions will have $B_T$ aligned in different directions 
but consider $B_T$ in one region of size $l_T$ 
in which $B_T$ is assumed constant.
Following (Larosa et al. 1996), 
define $\tau_r\equiv U_r(dU_r/dt)^{-1}=N\de t$ as the time scale for
the average {\it reflected} particle energy $U_r$ to double,
where $N$ is the number of required reflections and $\de t$ is the time
between reflections.  Since only a fraction $F$ are reflected, the
energization time averaged for {\it all} particles is then
$$\tau=U/dU/dt=(\de t/F)(N U/U_r),\eqno(8)$$
where $U$ is the average particle energy averaged over {\it all} particles.
we need to  estimate $N, \de t, F, U/U_r$.

To understand the role of $F$, 
two limits {\it not} usually distinguished  {\it must} 
be considered separately: For particles with speeds $v_{min}<v< V_{A}$, 
all reflections are ``head-on'' since the particles can never
catch up to the fluctuations which move at speeds along the
field lines  $\sim V_A$ (see figure 1).
For $v>>V_A  $, there are both ``catch-up'' and head-on reflections. 
Let  L1 and L2 label separate regimes where the average particle 
speed, $v_{ave}$, satisfies for L1: $v_{ave}>>V_A \sim v_{min}$
and L2: $v_{ave} \sim v_{min} \sim V_A$. 
That $v_{min}\sim V_A$ follows from the assumption that $\de B < B_T$ in (7).
L1 is appropriate for electrons
in a thermal equilibrium system whose magnetic pressure is
not dominant.  L2 is appropriate for protons in a 
thermal equilibrium system,  or for  both protons and electrons in
plasma of $\be_p\sim 1$ with the ratio of 
proton to electron temperature  $T_p/T_e \gsim 1000$---like ADAFs or HIT.
Define the corresponding energization times
$\tau_{L1}$ and $\tau_{L2}$ for the two limits.  
Limit L1 is the standard ``stochastic Fermi acceleration'' 
limit, and the energization time for L1
derived below simply, agrees with other treatments (e.g. 
Miller 1985, Melrose 1986).
The limit L2 produces a different formula.

To proceed further, I compute velocity moments of
reflected particles in the two cases L1 and L2 for
thermal and non-thermal distributions.  This is necessary for computing
$F$.  For a  power-law distribution,  
$dg_{nt}=(\la-1)(v/v_0)^{-\la}d(v/v_0),$
where $dg_{nt}/d(v/v_0)$ is the distribution function,  
$v_0$ is the lower cutoff on the power-law, and  $\la > 3$ will be assumed
(to avoid the appearance of logarithms).  
Integrating over the reflected particles gives
$$\int_{v_{min}}^\infty dg_{nt}=(v_{min}/v_0)^{1-\la},\eqno(9)$$
if $v_0< v_{min}$. If $v_0\ge v_{min}$, then $v_0$ replaces $v_{min}$ 
as the lower integral bound.
The average velocity of reflected particles is then 
$v_{r,ave}|_{nt} =(\la-1)v_0(v_{min}/v_0)^{2-\la}/(\la -2),$
when ${v_0 < v_{min}}$.
In the thermal case,
$dg_{th}= (4/\pi^{1/2})(v^2/v_{ave}^2)
Exp[-v^2/v_{ave}^2]d(v/v_{ave}),$ 
so that
$$\int_{v_{min}}^\infty dg_{th}=
1-Erf[v_{min}/v_{ave}]+(v_{min}/v_{ave})
Exp[-v^2_{min}/v_{ave}^2].\eqno(10)$$
I now use (9) and (10) to determine $F$, the fraction of particles that reflect.
Generally,  
$$F=\int_{v_{min}}^{\infty} f dg,\eqno(11)$$
where $f$ is the fraction of particles
that reflect at a given speed.
The $f$ is the ``area'' of the sphere (see Figure 1)
corresponding to particles which can be reflected, divided
by the total area, $4\pi v^2$: 
$$f=2\pi\int_{v_{{||}_-}}^{v_{{||}_+}} v_\perp[1+(dv_\perp/dv_{||})^2]^{1/2}dv_{||}/4\pi v^2=2\pi v [v_{||}]_{v_{{||}_-}}^{v_{{||}_+}}/4\pi v^2,\eqno(12)$$
where ${||}(\perp)$ indicates parallel (perpendicular) to $B_T$, and 
the second equality comes from using 
the equation for the circle centered at $V_A$ for the particle
speed, $v_\perp^2=v^2-(v_{||}-V_A)^2$.
The bounds $v_{{||}_\pm}$
are determined by finding the abscissa values at which the
line defining $\phi_{min}$ intersects (Figure 1) the circle defined
by $v_{ave}$.  Setting the equation for the lines, 
$v_\perp^2=v_{||}^2 tan^2\phi_{min}$, equal to that of the circle gives
$$v_{{||}_\pm} = V_A cos\phi'_{min}\pm cos\phi'_{min} 
(v_{ave}^2-V_A sin^2\phi'_{min})^{1/2}. \eqno(13)$$
For L1 (i.e. $v_{ave}>>V_A$), using (13) in (12) gives 
$f\sim cos\phi'_{min}\sim (\de B/B_T)^{1/2}
\sim 2 (2\pi v^2 cos\phi'_{min})/4\pi v^2=cos\phi'$.
This can be pulled out of the integral in (11).  Then because
$v_{min}/v_{ave}<<1$ in this limit,
$$F_{L1}=f=cos\phi_{min}',\eqno(14)$$ 
for both the non-thermal and
thermal cases.

For L2, ($v_{ave}\sim v_{min}$)
(12) and (13) give 
$f \lsim cos\phi'_{min}$.
This can be pulled out of the integral in (11).
For the non-thermal case, using (9) for the remaining integrand, 
I obtain  
$$F_{L2}\lsim cos\phi_{min}'[(v_{min}/v_{ave})(\la-1)/(\la-2)]^{1-\la}.
\eqno(15)$$
For the L2 thermal case, I use (10) instead of (9), noting that the
first two terms in (10) approximately cancel, giving
$$F_{L2} \sim cos\phi_{min}'(v_{min}/v_{ave})Exp[-v_{min}^2/v_{ave}^2].
\eqno(16)$$


Now consider $\de t=\de l/\lb|v_{||}|\rb$, the time between reflections,
where $\de l$ is the length scale of the fluctuation
$\de B$ and $\lb|v_{||}|\rb$ 
is the average magnitude of the reflected particles' velocity parallel to 
$B_T$. For L1, 
$\lb|v_{||}|\rb=v_{{||}_+}/2=
(v_{ave}/2) cos \phi'_{min}\sim(v_{ave}/2)(\de B/B_T)^{1/2}$ 
where the latter similarity follows for $\de B /B_T < 1$ in (7).
Thus 
$$\de t_{L1}= (2\de l/v_{ave})(B_T/\de B)^{1/2}.\eqno(17)$$
For L2 
$\lb|v_{||}|\rb\sim v_{{||}_+}/2\sim 
(V_A/2) cos^2\phi_{min}\sim(V_{w}/2)(\de B/B_T),$
so that 
$$\de t_{L2}= (2\de l/V_{w})(B_T/\de B).\eqno(18)$$

Consider now $N(U/U_r)$ appearing in (8).
For L1, $U/U_r\sim 1$, as the energy of reflected
particles is of order the average energy of all particles, 
but we must determine $N$.  For L1,
the energy gain is stochastic (Fermi 1949; Spitzer 1962; 
Eilek \& Hughes 1991; Larosa et al 1996) 
as the particles incur random walks through momentum space and 
$$N_{L1}=U^2/\lb\de U_+\rb^2,\eqno(19)$$
where $\lb\de U_+\rb$, is the average energy gain by a particle
from a head-on reflection.  
For L2, there are mainly head-on reflections,
so that $U_r/\lb\de U_+\rb \lsim N_{L2} \lsim U_r^2/\lb\de U_+\rb^2$. 
Since a lower bound on $\tau_r$ suffices, I 
employ $N_{L2} \gsim U_r/\lb\de U_+\rb$. This means that
For L2, 
$$[NU/U_r]_{L2}\gsim U/\lb\de U_+\rb.\eqno(20)$$

I now need $\lb\de U_+\rb$ for both L1 and L2.
The $\lb\de U_+\rb$ is determined by energy and momentum
conservation before and after a mirroring.
This gives 
$\lb\de U_+\rb\sim  2mV_A\lb|v_{||}|\rb$.  
For L1, using the value of $\lb |v_{||}|\rb$ calculated above then gives
$\lb\de U_+\rb\sim m V_A v_{ave}(\de B/B)^{1/2},$
and thus 
$$[NU/U_r]_{L1}=N_{L1}=(v_{ave}^2/4V_A^2)(B_T/\de B),\eqno(21)$$
while for L2, using the appropriate $\lb |v_{||}|\rb$ calculated
above (18) gives
$$[NU/U_r]_{L2}\gsim U/\lb\de U_+\rb=(v_{ave}^2/2 V_A^2)(B_T/\de B).\eqno(22)$$

Collecting the calculations of $N(U/U_r)$, $\de t$, and $F$ for L1 in (8) gives 
$$\tau_{L1}=(\de t_{L1}/F_{L1})[N U/U_r]_{L1}
=(\de l/4V_A) (v_{ave}/V_A)(B_T/\de B)^2=(l_T/4V_A) (v_{ave}/V_A)(\de l/l_T)^{1/2},\eqno(23)$$
where the last equality follows from assuming a
Kraichnan (1965) spectrum $(\de B/B)=(\de l/l_T)^{1/4}$ 
relating the magnetic to scale fluctuations.
Eq. (23) describes ``stochastic Fermi'' energization (Miller 1985; Melrose 1986).
Similarly for L2, using the appropriate above results
for $N(U/U_r)$, $\de t$, and $F$ in (8), I obtain 
$$\tau_{L2}=(\de t_{L2}/F_{L2})[N U/U_r]_{L2}
\gsim (\de l/V_A)(v_{ave}/V_A)^3(B_T/\de B)^{3/2}Exp[V_A^2/v_{ave}^2]$$
$$=(l_T/V_A)(v_{ave}/V_A)^3(\de l/l_T)^{5/8}Exp[V_A^2/v_{ave}^2],\eqno(24)$$
for the thermal case, while for the non-thermal case with  $v_{min}> v_0$, and $\la > 3$
$$\tau_{L2}\gsim (l_T/V_A)(v_{ave}/V_A)^{3-\la}(\de l/l_T)^{5/8},\eqno(25)$$
where the Kraichnan (1965) relation has again been used.

We see that each of the energy doubling times 
(23)-(25) depend only the particle average speeds $v_{ave}$ and not
on their particle mass.  But if electrons and protons have the same $v_{ave}$,
the protons have $(m_p/m_e)$ more energy. Thus each of (23)-(25)
shows that electrons take $(m_e/m_p)$ longer to drain the same amount
of energy.  When electrons and protons do not have the same $v_{ave}$
one population could be in L2 and the other in L1 and the comparison 
of energy doubling times becomes more subtle. This is 
because although in L1 there are many more reflections possible than in L2 
L1 has both energy gaining and 
energy losing reflections (i.e. note the smaller shaded area and absence
of symmetry in figure 1a compared to figure 1b). 
Thus the energization is second order as
expected for stochastic acceleration.  For L2 however, while there
are less reflections, they are mainly head-on (i.e. energy gaining).
These two effects (less reflections but mainly energy gaining=L1 
vs. more reflections but both energy gaining and energy losing =L2)
compete and the $\beta_p$ regime for which 
protons vs. electrons dominate the drain then also depends on 
the particle distribution of that population in the L2 limit.   
Another complication comes if the populations have the same $v_{ave}$ but
different distribution functions. Then one must compare (24) and (25).
We will study some of these cases more specifically in the next section.



\centerline{\bf 5. Application to Accretion Flows}

{\bf 5.1 Why Protons Are Likely Non-Thermal in ADAFs}

The discussion of section 4 is one
approach to  the mirroring or Fermi energization process.
Others include stochastic magnetic pumping (e.g. Hall \& Sturrock 1967) and
transit time pumping (e.g. Stix 1962, described as
the magnetic analogue of Landau damping). 
Achterberg (1981) showed that all small amplitude $(\de B << B_T)$
approaches in L1 to mirroring in a turbulent
plasma can also be described by quasi-linear diffusion of particles 
in momentum space, from  magnetosonic wave particle 
resonances at the Cherenkov 
resonance $(\omega_w-k_{||}v cos \phi)=0$.
The relevant waves have frequencies $\omega_w<<$ particle gyrofrequencies
(i.e. very long wavelengths compared to the gyro-radii) which
is equivalent to the adiabatic approximation discussed in section 4.

This resonance requires a minimum particle speed $v_{min}\sim V_A$ 
and also a minimum $sin \phi'$, as derived in section 4.  
The required minimum in $sin \phi'$ means that in order for 
particles to undergo repeated reflections and gain energy, their
momentum must be rapidly isotropized on a time scale shorter than
the time between reflections, which itself must be shorter
than the largest eddy turnover 
time.  The largest eddy turnover time is in turn shorter than
the ADAF infall time, given by $t_{in}\sim 1.8\ts 10^{-5}M r^{3/2}/\a$, 
where $M$ is the mass in units of $M_\odot$, 
and $r$ is the radius in Schwarzchild units.
But for ADAFs, Coulomb isotropization is not fast
enough:  the time scale for momentum isotropization from Coloumb collisions
is of order the time scale for thermalization and is given by (e.g. Spitzer 
1960; Mahadevan \& Quataert 1997)
$$t_{pp}=(2\pi)^{1/2}(n_p \sigma_T cln \La)^{-1}(m_p/m_e)^2(kT_p/m_p c^2)
\sim 10^{-2}\a(\be_{a}/0.5)^{3/2}M{\dot M},\eqno(26)$$
where $n_p$ is the proton number density, $\sigma_T$ is the Thomson
cross section, 
$ln \La$ is the Coulomb logarithm, and 
${\dot M}$ is the accretion rate is units of 
the Eddington value, $1.4\ts 10^{18} M$g/sec.
Setting (26) equal to $t_{in}$ 
shows that protons can only be Coulomb thermalized/isotropized  
well outside of the dominant energy emission location. i.e. for 
$r \gsim  100$  (Mahadevan \& Quataert 1997).  
(In fact, this feature is fundamental to enabling an ADAF solution.)
Thus the isotropization requires an additional kind of wave-particle
resonance.

Unlike the mirroring waves, the required isotropizing waves have wavelengths
of order the particle gyro-radius. Some, or all these small wavelength 
(Whistler, Alfv\'en or magnetosonic) 
waves can be generated by the particles themselves and then 
they do not transfer energy to the particles.  
Some fraction may also be generated directly from the turbulence 
in which case they can transfer energy to the particles.  This
latter possibility is explored in (e.g. Quataert \& Gruzinov 1998) 
as the primary means by which the turbulence dissipates into particles.
The resonances occur when the wave frequency in the
particle frame is an integer multiple of the particle gyrofrequency,
that is $\om  -k_{||}v cos \phi-N\Omega^*=0$,
where $\Om^*\equiv eBc/E_p$ and  $E_p$ is the total proton rest+kinetic 
energy.  Quataert \& Gruzinov (1998) show that Whistlers are not damped 
by protons. The short wavelength Alfv\'en 
waves $(\om=k_{||}V_A \lsim \Om_g \equiv eB/(m_i c)< \Om^*)$ are the most 
relevant for isotropization and have the approximate resonance condition
$-eBc/E_p -k_{||}v cos \phi=0$.  The condition 
$|cos\phi| <1$ then leads to the injection condition 
$E_p > (V_A/v) m_ic^2 (\Om_g/\om)$.  For $\om \sim \Om_g$, 
this leads to  $v_{min}\sim V_A $ for protons--
similar to the requirement of the mirroring waves derived earlier. 
  
So both types of resonant waves--long wavelength mirroring waves 
and short wavelength Alfv\'en waves (whether they accelerate or just
isotropize)--
have a proton minimum speed requirement of order $V_A$.  
The inefficiency of Coulomb thermalization,
and the need for wave particle resonances to dissipate the turbulence
means that a significant non-thermal particle population should be
produced.  
Since ADAFs are most commonly modeled with $\be_a\sim 0.5$,
non-thermal protons will likely be a fraction $\sim$O(1) of
the population.
The inefficiency of Coulomb collisions in ensuring a non-thermal population
 is fundamental.  Even when  stochastic Fermi energization (the L1 limit), 
can be shown to rigorously lead to a power-law distribution in the energized 
particles (e.g. Eilek \& Hughes 1991) 
efficient Coulomb scattering would thermalize the distribution.  The fact 
that Coulomb collisions are inefficient, as shown above, 
precludes redistribution of energy over the full population of protons.

Note that at least the isotropizing waves are also {\it implicitly} 
built into ADAFs because the standard models presume 
isotropic pressure and this would be impossible 
without wave-particle resonances. In fact, the plasma
must be ``collisional'' in the sense of wave-particle interactions,
even though it is ``collisionless'' with respect to Coulomb collisions.
In short, non-thermal protons should be a generic prediction of ADAFs.
This is consistent with observations (Mahadevan 1998), which
can distinguish between thermal and non-thermal proton distributions in an ADAF framework (and so far are not too sensitive to the proton power law index.)
  In principle, similar arguments could be applied to electrons with more
stringent resonance conditions.
However Mahadevan \& Quataert (1996) and Ghisselini et al. (1998)  
 argue that synchrotron self-absorption can 
thermalize weakly relativistic and non-relativistic  
electrons (at least those not produced from pion decay) 
under ADAF conditions during an in-fall time.
Thus I assume (25) applies to ADAF/HIT protons in the steady state,
and (24) to electrons. 

{\bf 5.3 Reflecting Waves, Scales of Dissipation,
and when Mirroring Preferentially Energizes Protons vs. Electrons}

Magnetosonic waves are the dominant mirrorers 
in the low amplitude limit $(\de B << B_T)$, as 
Alfv\'en waves are incompressible and compression is required for mirroring, 
though the relevant compression speed along the field lines 
is always $\sim V_A$ regardless of the wave mode.
Achterberg (1981) considered a magnetically dominated plasma at a single 
temperature, and focused only electrons.  Here we are interested in a two-temperature plasma and are considering both electrons and ions.
In general, both slow and fast waves may participate in the mirroring.  

Though magnetosonic waves dominate in the low
amplitude limit, this is not necessarily true in the large amplitude 
limit $(\de B\sim B_T)$.
Because the largest scales of turbulence in discs are the most compressive, 
the large amplitude limit is relevant when the 
the scale on which the mirroring can compete with the
cascade of energy from larger to smaller turbulent
scales is a large fraction of the outer turbulent scale $l_T$.
In this case, the energy could be compressively drained into particles
before it reaches smaller scales in the cascade where incompressible
modes of dissipation dominate.
Since large amplitude Alfv\'en waves are compressible 
(Alfv\'en \& Falthammar 1963), even they
could then contribute to the mirroring.  
Such Alfv\'en waves could even steepen to form shocks and perhaps shock-Fermi 
acceleration would be relevant.  This must be considered in future work.  
as will see that in fact the relevant mirroring scales can be large.

I now proceed to estimate the scales on which 
the favored particles are energized
and when protons vs. electrons are favored.
For low luminosity sources, ADAFs require $1-f\eta \le 0.01$, implying an
accretion efficiency $\le 1$\% of that for thin
discs (e.g. Rees et al. 1982; Narayan \& Yi 1995).  The respective
energization times then need to satisfy 
$$\tau_p/\tau_e\le \zeta \equiv T_p(1-f\eta)/T_e\lsim 10,\eqno(27)$$ 
where the subscript $p(e)$ indicates ions (electrons).
Since the turbulence cascades from large to small scales, I 
compare the scales of energy drain for protons, $(\de l)_p$, and
electrons, $(\de l)_e$, for which (27) is satisfied. 
The larger of the two length scales then  determines the dominant drain. 
Using (25) for ADAF protons and setting it equal to the eddy turnover time 
$l_T/V_A$ gives
$$(\de l)_p/l_T\sim (V_A/v_{p,ave})^{(24-8\la)/5},\eqno(28)$$ 
where
$v_{p,ave}$ is the average proton speed.  
For electrons, setting $\zeta$ times (24) equal to $l_T/V_A$ gives 
gives
$$(\de l)_e/l_T\sim \zeta^{-8/5}(V_A/v_{e,ave})^{24/5}Exp[-8V_A^2/5v_{e,ave}^2].\eqno(29)$$
Then we can see that  
$$(\de l)_p/(\de l)_e = \zeta^{8/5}k^{24/5}{\beta}_p^{4\la/5}Exp[8/(5 k^2 {\beta}_p)].\eqno(30)$$ 
This is  $>1$ for a range of parameters applicable to ADAFs
(e.g. $\beta_p \sim k\sim {\rm O}(1),\ \zeta\sim 10$). 
The same conclusion results when the particle distributions
are either {\it both} thermal or non-thermal.
{\it Protons can be favored to the required extent when compressive
modes dominate the turbulent dissipation.} 

Let us determine the scale on which the protons are dissipated.
From (28) we see that when $V_A\sim v_{ave,p}$, it is not hard
to have $(\de l)_p/l_T\sim {\rm O}(1)$, (recall $\la > 3$).  
This means that the compressive modes
may be very important and much of the energy in the turbulence
may drain before approaching the incompressible scales where 
Quataert \& Gruzinov (1998) is applicable.  In general,  
{\it the small amplitude limit may not be fully appropriate in describing
energy dissipation in ADAFs.}

Now consider a thermal plasma with $T_p=T_e$, which  
corresponds to radii outside the ADAF region
(Narayan \& Yi 1995) or to a thin, precursor disc.  
In this case, no matter which particles initially receive
the energy, Coulomb collisions redistribute this energy between electrons and
protons.  However, whether
protons vs. electrons receive the dissipated energy determines
the heating rate.
When $\beta_p\sim {\rm O}(1)$, the relevant limits
of interest are (23) for electrons and (24) for protons.
Using $v_{e,ave}=(m_p/m_e)^{1/2}v_{p,ave}$, the dissipation scale ratio becomes
$$(\de l)_p/(\de l)_e \sim (m_p/16m_e){ \beta}_p^{-7/5}Exp[-8/(5{\beta}_p)].\eqno(31)$$  
This is $< 1$ for ${\beta}_p\lsim  0.25$ and $>1$ 
for ${\beta}_p>   0.25$.
For ${\beta}_p >>1$, 
both electrons and protons are in the limit of (23), for which
$(\de l)_p/(\de l)_e\sim m_p/m_e$.
For $\beta_p < m_e/m_p$, both electrons and protons are in the limit 
of (24) for which  $(\de l)_p/(\de l)_e\sim (m_p/m_e)^{4/5}$.
In sum, electrons are favored only  for the range
$m_e/m_p \lsim  {\beta}_p \lsim 0.25$, while for 
${\be}_p$ outside this range protons are favored.  (The conditions of
low ${\beta}_p$ for which electrons are favored
may be found in solar flares (e.g. Larosa et al. 1996) 
and some thin accretion disc coronae models (Field \& Rogers 1993).



{\bf 5.4 Can Mirroring Help Form an ADAF?}

It is sometimes believed that purely a low enough accretion rate is enough 
to form an ADAF/HIT. However, unless the disc is already thick,
the critical accretion rate below which electrons and protons do not
couple by Coulomb collisions on an infall time as computed for a standard 
thin disc is far too low to be physically relevant.  
For a thin disc system to evolve into an ADAF, a mechanism is 
needed to form a thick disc first.  This may occur by thermal 
instability and mirroring may help.
The condition (e.g. Pringle 1981) for thermal instability is  
$$dln[q_t]/dT > dln[q_e^-]/dT.\eqno(32)$$  
If the instability proceeds from within an optically thick
disc, then we must compare blackbody emission to the 
heating. In the regime $\beta_p\lsim 0.25$ for the thermal disc, 
electrons are favored as shown in above, and (23) is applicable.
Taking the inverse of (23) for electrons, multiplying by $v_{ave}^2\propto T$
and differentiating gives $dln[q_t]/dT= 1/2T$.  
If the emission is blackbody, then 
$$dln[q_e^-]/dT=4/T,\eqno(33)$$
and the instability is not favored.
For $\beta_p \gsim 0.25$ protons are favored and using
(24) gives $dln[q_t]/dT= T^{-1}(1/\beta_p -1/2T)$ 
and still, even for $0.2 < \beta_p \lsim 0.5$ the thermal instability cannot
ensue. 

However, it is more likely that the formation of an ADAF
would proceed by thermal instability within the very surface
layer of the thin disc, and successive layers would 
eventually evaporate from the surface to form the thick ADAF disc.
The particle distribution in the very surface
layer could be non-thermal. To see how mirroring might help,
in the limit where the protons dominate the energy drain 
$(\be_p \gsim 0.2)$, I 
invoke (25) for protons, take its reciprocal, and multiply by
$v_{ave}^2 \propto T_p=T$ to obtain the quantity proportional to $q_t$. 
Then $dln[q_t]/dT=(\la-2)/2T$. 
For $\la > 8$ this can satisfy (32) when the emission is blackbody
(33).  For Bremsstrahlung $dln[q_e^-]/dT=3/2T$, and  (32) can be satisfied
when $\la > 5$.

\centerline{\bf  6. Conclusions}

Dissipation of turbulence in presumed ADAF sources must 
preferentially energize protons by a factor $\eta >99\%$ over electrons 
if  ADAFs are to account for the observed low luminosities.  
Compressive magnetic mirroring can in principle favor protons to the extent
required for a ratio of particle to magnetic pressure $\beta_p > 0.25$.
This is less stringent than the requirements of incompressible
modes which demand $\beta_p >>1$.
However, it is not 
easy to estimate what fraction of energy is dissipated in compressive
vs. incompressible modes.  An estimate of the scale at which
mirroring can compete with the transfer of energy down the turbulent
cascade seems to indicate that for $\beta_p \sim 1$, the scale can
be quite a large fraction of the outer turbulent scale---which for ADAFs/HIT
can be a large fraction of the disc size (Blackman 1998).
This has two implications:
1) a significant fraction of the energy may be dissipated in 
compressive modes and 2) the small amplitude approach to dissipation may 
not be valid. In the 
small amplitude limit, the relevant waves involved in the energy transfer
to particles are magnetosonic waves, as Alfv\'en
waves are not compressive and will not be damped by
mirroring.  However on the larger scales in the turbulent cascade,
the large amplitude limit is relevant.  Since large
amplitude Alfv\'en waves are compressive, they too may be involved
in compressive modes of dissipation.  
This presents additional complications for
future work.  Magnetic reconnection 
may also be a complication, however reconnection itself generates
turbulence, and possibly shock or direct acceleration processes which
may also favor protons, but 
it is important to know on what scale the reconnection is occurring.

Regardless of the fraction of energy dissipated in compressive
or incompressible modes, the fact ADAFs are ``Coulomb collisionless'' 
on the radial infall 
time scale seems to make a non-thermal proton population inevitable.  
The required dissipation of turbulence must proceed through wave particle
interactions, all of which act on only a subset of the particles.
Since ADAF models presume an isotropic pressure tensor, 
wave-particle resonances are implicitly assumed to play a role in ADAFs 
because Coulomb isotropization is necessarily too slow.  
The presence of a non-thermal proton population seems to indeed be
indicated by observations of the Galactic centre (Mahadevan 1998) when
modeled with an ADAF.  
A remaining fundamental problem which still 
needs more attention is  the question of a faster than Coulomb  
coupling between particles (Begelman \& Chiueh 1988) even if
the protons could receive the dissipated energy.

\ni Acknowledgments:  Thanks to M. Rees for stimulating discussions and 
insights.





\ni Achterberg, A., 1981, A\& A, 97, 259.

\ni Alfv\'en H \& Falthammar, C.-G., 1963, ``Cosmical Electrodynamics: Fundamental Principles'' (Clarendon:  Oxford)

\ni  Balbus, S.A. \& Hawley, J.F., 1991, ApJ, {\bf 376}, 214 

\ni Begelman M.C. \&  Chiueh,T., 1988, ApJ, {\bf 332}, 872 

\ni Bisnovatyi-Kogan G.S. \&  Lovelace, R.V.E., ApJ, 1997, {\bf 486}, L43


\ni Blackman, E.G., 1998, MNRAS in press.

\ni  Dreicer, H., 1959, Phys.Rev., {\bf 115}, 238 

\ni  Eilek, J.A., \& Hughes, P.A., 1991, in {\it Beams and Jets in
Astrophysics} ed. P.A. Hughes,  (Cambridge: Cambridge Univ. Press)

\ni Fabian, A.C. \& Rees, M.J., 1995, MNRAS, {\bf 277}, L55 

\ni  Fermi, E., 1949, Phys.Rev., {\bf 75}, 1169; Spitzer L., 1962, 
{\it Physics of Fully Ionized Gasses}, (New York:  Wiley)  

\ni Field G.B. \&  Rogers R.D., 1993, ApJ, $\bf 403$, 94 

\ni  Frank, J. King, A.R.  \& Raine, D.J., 1992,  {\it Accretion Power in 
Astrophysics}, (Cambridge:  Cambridge University Press)

\ni  Ghisselini, G.  Haardt, F. \& Svensson, R., 1998,  MNRAS in press. 

\ni Hall, D.E. \& Sturrock, P.A., 1967, Phys. Fluids, 10, 1593.

\ni  Holman, G.D., 1985, ApJ, {\bf 293} 584

\ni Ichimaru, S., 1977, Ap J, 214 840. 

\ni Kolmogorov, A., 1941, Dokl. Akad. Nauk SSR {\bf 30}, 299.

\ni  Kraichnan, R., 1965, Phys. Flu., {\bf 8}, 1385

\ni Kulsrud, R.M. \& Ferrari, A., 1971, Astrophys. Space Science, 12, 302.

\ni Larosa T.N. et al., 1996, ApJ, {\bf 467}, 454 

\ni Mahadevan, R. \& Quataert E., 1997, ApJ, 490, 605.

\ni  Mahadevan, R., 1998, Nature, in press.

\ni  Melrose, D.B., 1986, {\it Instabilities in Space and Laboratory Plasmas}, (Cambridge:  Cambridge Univ. Press)

\ni Miller, J.A., 1991, ApJ, {\bf 376} 342 

\ni  Narayan R. \& Yi, I. 1995, ApJ, {\bf 452}, 710 

\ni  Narayan,R. \&   Yi,I., 1995, \& R. Mahadevan, Nature {\bf 624}, 373 

\ni Narayan, R. \&   McClintock, J.E., \&  Yi, I., 1996, ApJ, {\bf 457}, 821 

\ni Narayan, R., Garcia,M.R., \&  McClintock J.E., 1997, ApJ, {\bf 478}, L79 

\ni  Paczy\`nski, B. \&  Bisnovatyi-Kogan, G.S., 1981, Acta Astron., {\bf 31}, 283 

\ni  Parker, E.N., 1979, {\it Cosmical Magnetic Fields}, (Oxford:
Clarendon Press)

\ni Pringle, J.E., 1981,  ARAA, {\bf 19}, 137

\ni Quataert, E., \&  Gruzinov, A., 1998, preprint astro-ph/9803112.

\ni Rees, M.J., 1982, AIP conference Proc on ``The Galactic Center'' 
(AIP: New York) 

\ni  Rees M.J., et al., 1982, Nature, {\bf 295}, 17 

\ni  Scudder J.D., et al., 1986, J. Geophs. Res. {\bf 91} 11053.

\ni  Shakura, N.I. \&  Sunyaev R.A., 1973, A.\&A., {\bf 24}, 337 

\ni Stix, T.H., 1962, {\it Theory of Plasma Waves} (New York: McGraw-Hill).

\ni Stone J.M. et al., 1996, ApJ, 463, 656.









\vfill
\eject

\centerline {\bf FIGURE 1 CAPTION}

Particle speed diagram for magnetic mirroring.  
The magnetic compression is assumed to move at
velocity $\sim -V_A$ along the field line 
and so an isotropically distributed population
of velocities in the lab frame has a spherical distribution centered around
$v_{||}'=V_A$.  The angle $\phi'_{min}$ and the speed $v_{min}$ bound
the respective minima needed for a particle 
to reflect at the magnetic compression. The weaker
the compression the larger these minima. The area inside the 
shaded region between the two circles represents
the particle speed region which can be reflected. Approximate 
schematics of the regimes 
a) L2 $(v_{ave}\sim V_A)$ and b) L1 $(v_{ave}>>V_A)$ of the
text are shown. L2 is relevant for $\beta_a\sim 1$ ADAFs. 
Note that only the region L1 strictly corresponds to stochastic energization 
since in this case the number of ``catch-up'' and head
on reflections are about equal whereas  region L2 has 
primarily ``head on'' (albeit less total) reflections.

\end